\documentclass[10pt,journal,compsoc,twoside]{IEEEtran}
\ifCLASSOPTIONcompsoc
  \usepackage[nocompress]{cite}
\else
  \usepackage{cite}
\fi
\ifCLASSINFOpdf
\else
\fi
\usepackage{setspace}
\usepackage[linesnumbered, ruled, vlined]{algorithm2e}
\usepackage{romannum}
\usepackage{subfigure} 
\usepackage{multirow}
\usepackage{bbding}
\usepackage{amsmath,amssymb}

\usepackage{scalerel}
\usepackage{tikz}
\usetikzlibrary{svg.path}
\definecolor{orcidlogocol}{HTML}{A6CE39}
\tikzset{
  orcidlogo/.pic={
    \fill[orcidlogocol] svg{M256,128c0,70.7-57.3,128-128,128C57.3,256,0,198.7,0,128C0,57.3,57.3,0,128,0C198.7,0,256,57.3,256,128z};
    \fill[white] svg{M86.3,186.2H70.9V79.1h15.4v48.4V186.2z}
                 svg{M108.9,79.1h41.6c39.6,0,57,28.3,57,53.6c0,27.5-21.5,53.6-56.8,53.6h-41.8V79.1z M124.3,172.4h24.5c34.9,0,42.9-26.5,42.9-39.7c0-21.5-13.7-39.7-43.7-39.7h-23.7V172.4z}
                 svg{M88.7,56.8c0,5.5-4.5,10.1-10.1,10.1c-5.6,0-10.1-4.6-10.1-10.1c0-5.6,4.5-10.1,10.1-10.1C84.2,46.7,88.7,51.3,88.7,56.8z};
  }
}
\newcommand\orcidicon[1]{\href{https://orcid.org/#1}{\mbox{\scalerel*{
\begin{tikzpicture}[yscale=-1,transform shape]
\pic{orcidlogo};
\end{tikzpicture}
}{|}}}}
\usepackage{hyperref} 

\begin{document}
%
\title{Stochastic Performance Analysis of Phase Decomposition in Hyperledger Fabric}

%

\author{Canhui~Wang\,,~\IEEEmembership{Member,~IEEE,}
        Xiaowen~Chu\,,~\IEEEmembership{Senior Member,~IEEE}
\IEEEcompsocitemizethanks{\IEEEcompsocthanksitem Canhui Wang is with the Computer Science and Technology Department, Huaqiao University, Xiamen, China.\protect\\
E-mail: chwang@hqu.edu.cn

\IEEEcompsocthanksitem Xiaowen Chu is with the Data Science and Analytics Thrust, The Hong Kong University of Science and Technology (Guangzhou), China.\protect\\
E-mail: xwchu@ust.hk

}
}

\IEEEtitleabstractindextext{%
\begin{abstract}
Hyperledger Fabric is one of the most popular permissioned blockchain platforms. Although many existing works on the overall system performance of Hyperledger Fabric are available, a decomposition of each phase in Hyperledger Fabric remains to be explored. Admittedly, the overall system performance of Hyperledger Fabric might provide an end-user with satisfied performance information when invoking a transaction; however, it is far from informative when deploying a distributed system with specific performance goals, except for understanding each phase in Hyperledger Fabric. Besides, it is challenging to study the performance of a distributed system with many dependent phases like Hyperledger Fabric, where the output of a phase becomes the input of the next phase, and each phase’s output might not follow a Poisson distribution. In this paper, we develop a measurement framework to characterize each phase’s transaction and block data in Hyperledger Fabric based on the Fabric SDK Nodejs, where we thoroughly analyze and open-source the implementation details of the measurement framework. We evaluate the performance of Hyperledger Fabric and have some interesting observations, e.g.,~1) The number of CPU cores has a linear impact on the throughput of an endorsing peer.~2) The Raft-based ordering service shows good scalability with the number of ordering service nodes.~3) The communication latencies between the client and service in Hyperledger Fabric are significant. We then identify each phase's dominant latency in Hyperledger Fabric via primitive operation analysis and propose a stochastic computation model for performance analysis. Specifically, an endorsing peer in the execute phase is modeled as a $M/D/c$ queue, the OSN leader in the order phase is modeled as a $G/G/1$ queue, and a committing peer in the validate phase is modeled as a $G/G/1$ queue. We also use the alpha-beta communication model to analyze the corresponding communication latency. Finally, we validate the accuracy of the performance model on both local and cloud clusters. The experiment results and the performance model help guide the deployment of the Hyperledger Fabric service.
\end{abstract}

\begin{IEEEkeywords}
Hyperledger fabric, blockchain, phase decomposition, stochastic computation model, alpha-beta communication model
\end{IEEEkeywords}}

\maketitle

\IEEEdisplaynontitleabstractindextext

\IEEEpeerreviewmaketitle

\IEEEraisesectionheading{\section{Introduction}\label{sec:introduction}}
%
\IEEEPARstart{B}{lockchain} is a Distributed Ledger Technology (DLT) originally proposed to solve the double-spending attack in the Bitcoin (BTC) cryptocurrency \cite{nakamoto2008bitcoin}. A blockchain is a chain of blocks replicated among participants on a peer-to-peer (P2P) network. An interesting property \cite{wang2022dissecting} of blockchain is that it does not rely on any third party for transaction clearing; instead, it stores and updates all valid historical transactions for future reference and hence can determine whether a new coming transaction is valid. People categorize blockchain into permissioned blockchain \cite{gai2019permissioned,  putra2021trust, helliar2020permissionless, bakos2021permissioned} and permissionless blockchain \cite{peng2021privacy}. Specifically, a permissioned blockchain requires prior authorization with which only authenticated users can access specific blockchain data, while a permissionless blockchain does not require that. The permissioned blockchain is prevalent in many fields of the industry \cite{de2020survey, lao2020survey} where data security is a top concern, e.g., healthcare \cite{lakhan2022blockchain, peng2023patient} and e-commerce.

Hyperledger Fabric, a popular permissioned blockchain, is a distributed system with multiple components, including peer, ordering service, and smart contract. Each transaction proposed to update the global state of the blockchain database goes through these components chronologically, except for the read-only transaction that does not update the global state. To reduce component dependency, the recent versions of Hyperledger Fabric, i.e.,~v1.0+, apply a loosely coupled architecture called the execute-order-validate model \cite{androulaki2018hyperledger}. It encapsulates these components into various pluggable modules and then decouples the whole system into three phases: execute, order, and validate, where the output of a phase becomes the input of the next phase. Specifically, the execute phase receives new proposals from the client and simulates new transactions for the client. The order phase puts these new transactions from the client into a new block and then forwards the new block to the validate phase. Finally, the validate phase validates each transaction, commits the new block, and replies to the client.

Performance modeling is important to understand the Hyperledger Fabric system. Our motivations for performance modeling and analysis of Hyperledger Fabric are as follows. First, many existing works studied the system performance of Hyperledger Fabric from the perspective of an overall system \cite{klenik2022porting, jeong2021multilateral, lohachab2021performance, fan2021performance, hang2021optimal} and few explored each phase. Although an overall system performance may provide an end-user with enough performance information, it is far from informative for a system developer to deploy the Hyperledger Fabric system to meet specific performance goals, except for understanding each phase in Hyperledger Fabric. Besides, it still lacks a measurement method for each phase decomposition and a comprehensive measurement framework. Therefore, our motivation is to propose a measurement framework to decompose each phase.

Second, performance modeling of the Raft-based ordering service is important to Hyperledger Fabric~2.2~LTS because the Raft-based ordering service is the only choice to handle the consensus challenge among blockchains in the recent version of Hyperledger Fabric~2.2~LTS, and other alternatives like Kafka and Solo have been deprecated since Fabric~v2.x. Although a few recent empirical studies \cite{wang2020performance, barger2021byzantine} on the scalability of the Raft-based ordering service are available, there are few performance models of the scalability of the Raft-based ordering service. Admittedly, there are a few performance models of the Raft-based ordering service; but they did not consider the scalability of the ordering service, and the corresponding ordering services are in a solo mode with a single ordering service node (OSN) \cite{xu2021latency, yuan2020performance, meng2021consortium}. Therefore, our motivation is to study and model the scalability of the Raft-based ordering service.

Third, few related works study the dominant resource of each phase while existing works focus on the impact of various resources on the overall system performance \cite{jeong2021multilateral, desai2021blockfla, gai2022blockchain, fotia2022trust, neha2022systematic}. Identifying each phase's dominant resource is essential when deploying Hyperledger Fabric service. Therefore, our motivation is to identify each phase's dominant resource. Specifically, we identify the impact of CPU cores on an endorsing peer in the execute phase, the impact of ordering service nodes~(OSNs) on the ordering service in the order phase, and the impact of disk IOs on a committing peer in the validate phase via primitive operation analysis. We then propose a stochastic computation model and an alpha-beta communication model for each phase analysis and validate the proposed model in both local and cloud clusters. Finally, the main contributions of this paper can be summarized as follows,

\begin{itemize}
    \item We propose a measurement framework to decompose the Hyperledger Fabric service and study each phase's transaction and block characteristics. We make the measurement framework and numerical results publicly available to the research community for recent Hyperledger Fabric 2.2~LTS. 

    \item We study the scalability of the Raft-based ordering service with the number of ordering service nodes and propose an alpha-beta communication model for the scalability of the Raft-based ordering service. We validate the communication model in local and cloud networks.

    \item We propose a stochastic computation model to analyze the performance of each phase in Hyperledger Fabric, where an endorsing peer with $c$ CPU cores in the execute phase is modeled as a $M/D/c$ queue, the process of an OSN leader in the order phase is modeled as a $G/G/1$ queue, and a committing peer in the validate phase is modeled as a $G/G/1$ queue. We validate the computation model in local and cloud clusters.
\end{itemize}

The remainder of this paper is organized as follows. Section~2 discusses related work in performance measurement and modeling of Hyperledger Fabric. Section~3 identifies each phase's dominant resource via primitive operation analysis. Section~4 presents a stochastic computation model for each phase in Hyperledger Fabric and an alpha-beta communication model for the scalability of the Raft-based ordering service. The design and implementation of the measurement framework are analyzed in Section~5, and the validation results are shown in Section~6. Finally, Section~7 concludes this paper.

\section{Related Work}
Many existing works \cite{berendea2020fair, wang2020performance, gorenflo2020fastfabric, sharma2019blurring, nasir2018performance, nakaike2020hyperledger, nguyen2019impact, kuzlu2019performance, thakkar2018performance} studied the overall system performance of Hyperledger Fabric. For example, Nakaike \textit{et al.} \cite{nakaike2020hyperledger} explored the impacts of database systems on the system performance of Hyperledger Fabric. They ran a Hyperledger Fabric GoLevelDB (HLF-GLDB) benchmark and found that the data compression of GoLevelDB could bring a significant performance issue to the system performance of Hyperledger Fabric. Nguyen \textit{et al.} \cite{nguyen2019impact} investigated the impacts of network delay on Hyperledger Fabric. They deployed the Hyperledger Fabric over an area network between France and Germany to observe the effects of network delays on the performance of Hyperledger Fabric and found that Hyperledger Fabric does not provide sufficient consistency guarantees to be deployed in critical network environments. Thakkar \textit{et al.} \cite{thakkar2021scaling} studied the scalability of the validate phase in Hyperledger Fabric. They found that the validate phase is a performance bottleneck in Hyperledger Fabric and proposed an improvement scheme called sparse peer to improve the validation phase's scalability via validating only a subset of transactions. They \cite{thakkar2018performance} also studied the impact of various system parameters such as batch size, channels, and endorsement policy on the performance of Hyperledger Fabric. They found that the validation phase, such as endorsement policy verification, policy validation of transactions, state validation, and commitment, brings performance issues to Hyperledger Fabric. Most existing works focus on the overall system performance of Hyperledger Fabric. Although the overall system performance may provide an end-user with enough performance information when invoking or querying a transaction, it is far from informative for a system deployer to deploy the Hyperledger Fabric service on a local or cloud cluster to meet specific performance goals except for understanding each phase in Hyperledger Fabric. However, quite a few related works studied the performance characteristics of each phase, and a comprehensive measurement method for each phase in Hyperledger Fabric is still lacking. Therefore, we follow the classic execute-order-validate model and decompose the overall system into three phases. We then propose a measurement method to study each phase.

Performance modeling is important to identify the dominant resource in Hyperledger Fabric. It helps to improve the performance of the Hyperledger Fabric service quantitatively. The ordering service is one of the core components of Hyperledger Fabric, and there are some existing works \cite{sukhwani2018performance, xu2021latency, yuan2020performance, meng2021consortium} studied the performance of the ordering service. For example, Xu \textit{et al.} \cite{xu2021latency} provided a queue model to calculate the transaction latency under various network configurations, such as batch size and block interval, and proposed a $M/M/1$ queue to model the PBFT-based ordering service in a solo mode. Yuan \textit{et al.} \cite{yuan2020performance} analyzed the performance of the ordering service under different system configurations and proposed a Petri Net-based model to study the performance of ordering service in a solo mode. Meng \textit{et al.} \cite{meng2021consortium} analyzed the security properties of the ordering service and the impact of delaying endorsement messages in consortium blockchain protocols. However, the recent version of Hyperledger Fabric 2.2~LTS only supports the Raft-based ordering service, while other alternatives like Kafka and Solo-based ordering services have been deprecated since Fabric v2.x. Also, quite a few related works studied the scalability of the Raft-based ordering service in Hyperledger Fabric, especially when the number of ordering service nodes is more than one \cite{sukhwani2018performance, xu2021latency, yuan2020performance, meng2021consortium}. Therefore, we focus on communication modeling of the scalability in the Raft-based ordering service. Moreover, we propose a stochastic computation model to study the performance of each phase. We validate the proposed performance models in local and cloud clusters.

\section{System Description}
This section will follow the classic execute-order-validate model to decompose the overall system into three phases and identify each phase's dominant resource. 


\subsection{The Execute Phase}

\begin{figure}[t]
  \centering
  \includegraphics[scale=0.70]{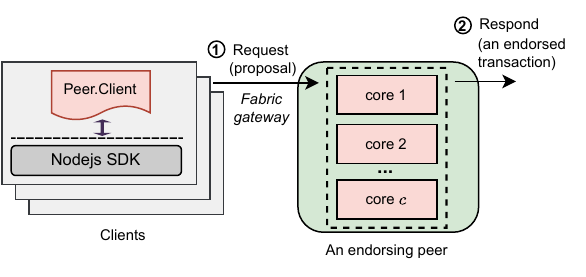}
  \caption{An illustration of the dominant resource in the execute phase, where an endorsing peer has $c$~CPU cores. In step~\textcircled{1}, the client initiates a transaction proposal and sends the proposal to the endorsing peer. In step~\textcircled{2}, the endorsing peer simulates, endorses, and passes back the response to the client.}  
\end{figure}

\textit{Dominant Resource in the Execute Phase.} The primitive operations of the execute phase involve two parts: proposal verification and transaction simulation. Specifically, upon receiving a transaction proposal, the execute phase completes proposal verification (e.g., the transaction proposal is well formed, it has not been submitted before, the client’s signature is valid, and the client is authorized to submit a transaction proposal). After proposal verification, the execute phase completes the transaction simulation (i.e., it takes the transaction proposal as an input argument to invoke the user chaincode and get the transaction results, and then, the execute phase uses an elliptic curve digital signature algorithm (ECDSA-256) to make a digital signature and return the transaction results and the signature to the client). Fig.~1 illustrates the dominant resource in the execute phase. In step~\textcircled{1}, the client initializes a transaction proposal and submits the transaction proposal to an endorsing peer via fabric gateway in the execute phase. In step~\textcircled{2}, the endorsing peer simulates and endorses the proposal and passes back the response to the client. Note that the execute phase handles these primitive operations concurrently on multiple CPU cores. It implies that the service times of transactions can be parallelized in the execute phase. 

\subsection{The Order Phase}

\begin{figure}[t]
  \centering
  \includegraphics[scale=0.66]{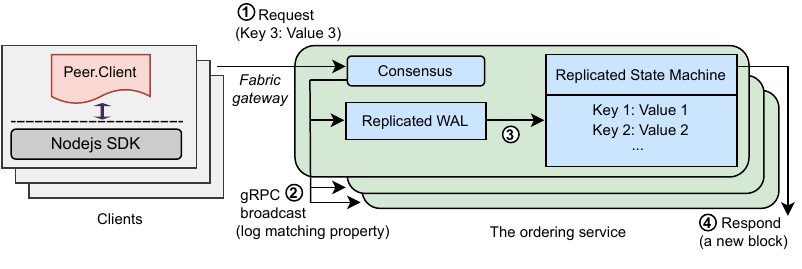}
  \caption{An illustration of the dominant resource in the order phase, where an ordering service has $k=3$~OSNs, i.e., one OSN leader and two OSN followers. In step~\textcircled{1}, the client requests to write an endorsed key~3. In step~\textcircled{2}, the consensus module receives the request from the client, encapsulates the request into a log entry, and replicates the log entry to the replicated write-ahead log~(WAL) of OSN followers. To ensure that the replicated state machines of OSNs have the same sequenced input, the Raft consensus algorithm enforces a log-matching property to guarantee consistent log entry sequences among OSNs. In step~\textcircled{3}, upon receiving the replies of the sequenced log entry from the OSN followers, the OSN leader feeds the sequenced log entries to the replicated state machine to apply the state of the key and then cuts a new block in step~\textcircled{4}.}
\end{figure}

\textit{Dominant Resource in the Order Phase.} The primitive operations of the order phase involve two parts: the communication latency between the OSN leader and the OSN followers in the Raft protocol and the idle time when batching a new block. Specifically, from the perspective of computing resources, the transaction flow in the order phase is as follows: in step~\textcircled{1}, the client submits transactions to the local memory of the OSN leader over the network. In step~\textcircled{2}, the OSN leader concurrently broadcasts these new transactions to the local memory of the OSN followers over the network while enduring that at least a majority of OSN followers maintain the same sequence of these new transactions. In steps~\textcircled{3},~\textcircled{4}, the OSN leader puts the transactions from a preordered sequence into a new block, commits the new block to local memory storage instead of disk storage, replies to peers, and forwards the new block to the OSN followers, as illustrated in Fig.~2. Remark that a block is technically a tiny offset pointer since most OSN followers have already successfully maintained a preordered sequence of new transactions. 


The second primitive operation of the order phase is the block cutter on the OSN leader. The Raft consensus protocol handles data transaction by transaction, while the Raft-based ordering service handles data block by block. In other words, a transaction that quickly achieves a Raft consensus may not be put into a new block immediately due to different granularity levels. Therefore, it needs to wait until enough transactions arrive and a new block is created. Hence the idle time of a transaction when batching a new block is introduced by the block cutter on the OSN leader. Two key configuration parameters determine the idle time: \textit{BatchTimeout} and \textit{BatchSize}. The \textit{BatchTimeout} determines the time to wait until creating a new block, and the \textit{BatchSize} determines the total number of transactions and the total transaction size of a new block. The ordering service generates a new block if either configuration parameter is satisfied.  

\subsection{The Validate Phase}

\begin{figure}[t]
  \centering
  \includegraphics[scale=0.66]{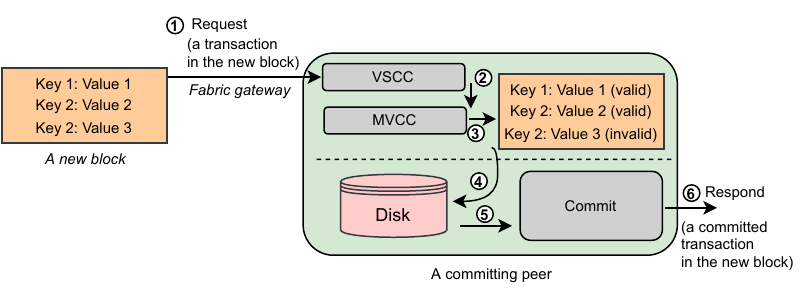}
  \caption{An illustration of the dominant resource in the validate phase where there is a committing peer. Consider that the ordering service creates a new block with three transactions. The first transaction writes key 1 with value 1, the second one writes key 2 with value 2, and the third one writes key 2 with value 3. In step~\textcircled{1}, the block is transmitted to the peer in the validate phase. Upon receiving a new block, the peer in step~\textcircled{2} validates each transaction within the block and ensures that the endorsement policy is fulfilled. Then, the multi-version concurrency control (MVCC) chaincode handles concurrent transactions in step~\textcircled{3}. Concurrent transactions with read-write conflicts are marked as invalid. Both valid and invalid transaction data are stored in the blockchain on the local database. In contrast, the state of an invalid transaction is not applied to the global state of the Hyperledger Fabric blockchain. In step~\textcircled{4} and~\textcircled{5}, the committer stores block data to the local database, and applies the state of each valid transaction to the global state. In step~\textcircled{6}, the committer notifies the client about the valid or invalid state status.}
\end{figure}

\textit{Dominant Resource in the Validate Phase.} The primitive operations of the validate phase involve two parts: the validate operation and the commit operation. As shown in Fig.~3, there are three transactions within a new block in step~\textcircled{1}. Each transaction in a new block will be validated to ensure its endorsement policy is fulfilled in step~\textcircled{2} and the read set matches the same key in the global state in step~\textcircled{3}, meaning that the referred preceding valid transaction has been committed successfully. In other words, Hyperledger Fabric applies the validation of the read set to prevent the double-spending attack. In steps~\textcircled{4},~\textcircled{5}, and~\textcircled{6}, valid and invalid transaction data are stored in the blockchain on the local disk, and only valid transaction data are applied to the global state of the Hyperledger Fabric blockchain.

\section{System Modeling}
This section will propose a stochastic computation model and an alpha-beta communication model; particularly, we will study the performance of each phase and then combine all three phases. Table~1 summarizes the key variables of the system model, where variables are deterministic except for random variables.

\begin{table*}[t]
\footnotesize 
	\centering
	\caption{Key Variables}
	\scalebox{0.96}{
\begin{tabular}{|c|l|l|}
\hline
Variable         & Description                                                                               & Units         \\ \hline \hline 
$\zeta$      & The number of clients                                                                     & Dimensionless \\ \hline
$c$          & The number of CPU cores of an endorsing peer in the execute phase                         & Dimensionless \\ \hline
$k$          & The number of OSNs in the order phase where $k$ is an odd number                                   & Dimensionless \\ \hline
$CV$          & The coefficient of variation                                                                             & Dimensionless          \\ \hline
$m$          & The transaction size, i.e., 3~KB                                                                          & KB          \\ \hline
$\alpha$     & The constant communication overhead, where it takes $\alpha=10$ ms in a cluster                                                               & Millisecond   \\ \hline
$\beta$      & The effective network bandwidth                                                                  & Mbps        \\ \hline \hline
$\lambda^e$  & The transaction arrival rate of an endorsing peer with $c$ CPU cores in the execute phase & TPS           \\ \hline
$\mu^e$      & The transaction service rate of an endorsing peer with $c$ CPU cores in the execute phase & TPS           \\ \hline
$\rho^e$     & The utilization of an endorsing peer with $c$ CPU cores in the execute phase              & Dimensionless \\ \hline
$\psi^e$     & The throughput of an endorsing peer with $c$ CPU cores in the execute phase               & TPS           \\ \hline
$T^e=\{\tau_i^e,i=1,2,\}$        & The $\tau_i^e$ is the overall latency of the $i$-th transaction in the execute phase                            & Second        \\ \hline
$T_{c2e}^e=\{\tau_{i, c2e}^e,i=1,2,\}$ & The $\tau_{i, c2e}^e$ is the communication latency of the $i$-th transaction from client to endorsing peer                     & Second        \\ \hline
$T_{e2c}^e=\{\tau_{i,e2c}^e,i=1,2,\}$ & The $\tau_{i,e2c}^e$ is the communication latency of the $i$-th transaction from endorsing peer to client                     & Second        \\ \hline
$T_{comm}^e=\{\tau_{i,comm}^e,i=1,2,\}$ & The $\tau_{i,comm}^e$ is the communication latency of the $i$-th transaction in the execute phase                     & Second        \\ \hline
$T_{s}^e=\{\tau_{i,s}^e,i=1,2,\}$    & The $\tau_{i,s}^e$ is the service time of the $i$-th transaction in the execute phase                            & Second        \\ \hline
$T_{q}^e=\{\tau_{i,q}^e,i=1,2,\}$      & The $\tau_{i,q}^e$ is the queue latency of the $i$-th transaction in the execute phase                             & Second        \\ \hline \hline
$\lambda^r$  & The transaction arrival rate of the OSN leader in the order phase                         & TPS           \\ \hline
$\mu^r$      & The transaction service rate of the ordering service with $k$ OSNs in the order phase      & TPS           \\ \hline
$\rho^r$     & The utilization of the ordering service with $k$ OSNs in the order phase                  & Dimensionless \\ \hline
$\psi^r$     & The throughput of the ordering service with $k$ OSNs in the order phase                   & TPS           \\ \hline
$T^r=\{\tau_i^r,i=1,2,\}$        & The $\tau_i^r$ is the overall latency of the $i$-th transaction in the order phase                             & Second        \\ \hline
$T_{c2l}^r=\{\tau_{i,c2l}^r,i=1,2,\}$ & The $\tau_{i,c2l}^r$ is the communication latency of the $i$-th transaction from client and to OSN leader                       & Second        \\ \hline
$T_{l2f}^r=\{\tau_{i,l2f}^r,i=1,2,\}$ & The $\tau_{i,l2f}^r$ is the communication latency of the $i$-th transaction from OSN leader to followers                       & Second        \\ \hline
$T_{comm}^r=\{\tau_{i,comm}^r,i=1,2,\}$ & The $\tau_{i,comm}^r$ is the communication latency of the $i$-th transaction in the order phase                       & Second        \\ \hline
$T_{s}^r=\{\tau_{i,s}^r,i=1,2,\}$      & The $\tau_{i,s}^r$ is the service time of the $i$-th transaction in the order phase                             & Second        \\ \hline
$T_{idle}^r=\{\tau_{i, idle}^r,i=1,2,\}$      & The $\tau_{i, idle}^r$ is the idle time of the $i$-th transaction when batching a block in the order phase                             & Second        \\ \hline
$T_{q}^r=\{\tau_{i,q}^r,i=1,2,\}$      & The $\tau_{i,q}^r$ is the queue latency of the $i$-th transaction in the order phase                               & Second        \\ \hline \hline
$\lambda^v$  & The transaction arrival rate of a committing peer in the validate phase                   & TPS           \\ \hline
$\mu^v$      & The transaction service rate of a committing peer in the validate phase                     & TPS           \\ \hline
$\rho^v$     & The utilization of a committing peer in the validate phase                                & Dimensionless \\ \hline
$\psi^v$     & The throughput of a committing peer in the validate phase                                 & TPS           \\ \hline
$T^v=\{\tau_i^v,i=1,2,\}$        & The $\tau_i^v$ is the overall latency of the $i$-th transaction in the validate phase                          & Second        \\ \hline
$T_{comm}^v=\{\tau_{i,comm}^v,i=1,2,\}$ & The $\tau_{i,comm}^v$ is the communication latency of the $i$-th transaction in the validate phase                    & Second        \\ \hline
$T_{s}^v=\{\tau_{i,s}^v,i=1,2,\}$      & The $\tau_{i,s}^v$ is the service time of the $i$-th transaction in the validate phase                         & Second        \\ \hline
$T_{q}^v=\{\tau_{i,q}^v,i=1,2,\}$      & The $\tau_{i,q}^v$ is the queue latency of the $i$-th transaction in the validate phase                            & Second        \\ \hline
\end{tabular}
}
\end{table*}

\subsection{The Execute Phase}

\textit{Steady-State Throughput of an Endorsing Peer.} Assume that the transaction arrival from a client follows a Poisson distribution, the service time is deterministic, and an endorsing peer serves $\zeta$ clients. Let the transaction arrival rate of the client $i$ be $\lambda_i^e$ transactions per second. Therefore, the execute phase has a transaction arrival rate $\lambda^e=\sum_{i=1}^{\zeta}\lambda_i^e=\zeta \lambda_i^e$, where $\lambda_i^e$ is the transaction arrival from the client $i$ to the target endorsing peer. And due to the merging property of Poisson processes, the merged arrival of the execute phase also follows a Poisson distribution.

We model an endorsing peer with $c$ CPU cores as a $M/D/c$ queue because the execute phase's endorsing operations can be parallelized. Therefore, we have a linear model of the steady-state throughput of an endorsing peer with $c$ CPU cores, where the service rate of the CPU core $i$ of the endorsing peer is denoted by $\psi_i^e$ and $\rho^e=\frac{\lambda^e}{c\mu^e}$, as follows, 

\begin{equation}
    \psi^e=\sum_{i=1}^{c}\psi_i^e=c\psi_i^e; \psi_i^e=\mu^e\rho^e
\end{equation}

\textit{Expected Latency in the Execute Phase.} Consider that the overall latency of a transaction in the execute phase is a random variable $T^e=\{\tau_i^e,i\in \mathbb{Z}^+\}$ where $\tau_i^e$ is the overall latency of the $i$-th transaction in the execute phase. $T^e$ consists of three parts: the communication latency of a transaction in the execute phase denoted by a random variable $T_{comm}^e=\{\tau_{i,comm}^e,i\in \mathbb{Z}^+\}$ where $\tau_{i,comm}^e$ is the communication latency of the $i$-th transaction in the execute phase, the service time of a transaction in the execute phase denoted by a random variable $T_{s}^e=\{\tau_{i,s}^e,i\in \mathbb{Z}^+\}$ where $\tau_{i,s}^e$ is the service time of the $i$-th transaction in the execute phase, the queue latency of a transaction in the execute phase denoted by a random variable $T_{q}^e=\{\tau_{i,q}^e,i\in \mathbb{Z}^+\}$ where $\tau_{i,q}^e$ is the queue latency of the $i$-th transaction in the execute phase. Therefore, we have the expected latency of a transaction in the execute phase, where random variables $T^e=T_{comm}^e+T_{s}^e+T_{q}^e$, as follows, 

\begin{equation}
    \centering
    \begin{split}
        \mathop{\mathbb{E}}\left [ T^e \right ]=\mathop{\mathbb{E}}\left [ T_{comm}^e \right ]+\mathop{\mathbb{E}}\left [ T_{s}^e \right ]+\mathop{\mathbb{E}}\left [ T_{q}^e \right ]\;\;\;\;\;\;\;\;\;\;\\
        =2\alpha + \frac{2m\lambda^e}{\beta^e}+\frac{1}{\mu^e}+\frac{1}{2}(1+f(c)g(\rho^e))\mathop{\mathbb{E}}\left [ T_q^{M/M/c} \right ]
    \end{split}
\end{equation}

\noindent where, transactions between the client and the endorsing peer are transferred over the network in pipeline parallelism. Thus, we model the expected communication latency from the client to the endorsing peer as $\mathop{\mathbb{E}}\left [ T_{c2e}^e \right ]=\alpha+\frac{m'\lambda^e}{\beta_{c2e}^e}$, and the expected communication latency from the endorsing peer to the client as $\mathop{\mathbb{E}}\left [ T_{e2c}^e \right ]=\alpha+\frac{m\lambda^e}{\beta_{e2c}^e}$, where $\lambda^e$ is the transaction arrival rate, $m$ is the transaction size, $m'$ is the transaction proposal size, $\beta_{c2e}^e$ is the effective bandwidth from the client to the endorsing peer, and $\beta_{e2c}^e$ is the effective bandwidth from the endorsing peer to the client. In practice, the transaction size (i.e., around 3KB) minus the size of a 256-bit ECDSA signature equals the transaction proposal size, and we approximate $m=m'$. Moreover, the bi-directional communication between the client and the endorsing peer (i.e., from the client to the endorsing peer, from the endorsing peer to the client) uses the same gRPC communication protocol. Therefore, we approximate the expected communication latency of a transaction in the execute phase as $\mathop{\mathbb{E}}\left [ T_{comm}^e \right ]=\mathop{\mathbb{E}}\left [ T_{c2e}^e \right ]+\mathop{\mathbb{E}}\left [ T_{e2c}^e \right ]=2\alpha+\frac{2m\lambda^e}{\beta^e}$, where $\alpha$ is a constant communication overhead, and $\beta^e$ is the effective bandwidth between the endorsing peer and the clients in the execute phase. Second, the expected service time of the execute phase can be modeled as $\mathop{\mathbb{E}} \left [ T_s^e \right ]=\frac{1}{\mu^e}$, where $\mu^e=max(\psi_i^e), i \in \left [ 1,c \right ]$ is the maximum throughput of a single CPU core in the execute phase. Third, the queue latency of the execute phase is modeled as a $M/D/c$ queue. We approximate \cite{yan2018efficient, di2022performability} the $M/D/c$ queue because there is no closed-form solution of the $M/D/c$ queue \cite{zhang2019mark}. Therefore, the expected queue latency of the execute phase can be modeled as $\mathop{\mathbb{E}} \left [ T_q^e \right ]= \mathop{\mathbb{E}} \left [ T_q^{M/D/c} \right ]=\frac{1}{2}(1+f(c)g(\rho))\mathop{\mathbb{E}}\left [ T_q^{M/M/c} \right ]$, where $f(c)=\frac{\left ( c-1 \right )\left ( \sqrt{4+5c} -2\right )}{16c}$ and $g(\rho^e)=\frac{1-\rho^e}{\rho^e}$.

\subsection{The Order Phase}

\textit{Steady-State Throughput of an OSN Leader.} Assume that the transaction arrival of the order phase follows a General distribution and the service time also follows a General distribution. We have the transaction arrival rate of the order phase $\lambda^r=\psi^e$ because the output of the execute phase becomes the input of the order phase. Let's consider the process of transaction ordering in the OSN leader. Denote the throughput of the ordering service with $k$ OSNs by $\psi^r$. We model the process of transaction ordering in the OSN leader as a $G/G/1$ queue because the job of transaction ordering is to sequence transactions from the clients, meaning that the service times of sequencing transactions in the OSN leader are not parallelized. Therefore, we have the steady-state throughput of the OSN leader with $i$ OSN followers, where $\rho^r=\frac{\lambda^r}{\mu^r}$, as follows,

\begin{equation}
    \psi^r=\psi_i^r,i=3,5,7\cdots,k-1; \psi^r=\mu^r \rho^r
\end{equation}

\textit{Expected Latency in the Order Phase.} Consider that the overall latency of a transaction in the order phase is a random variable $T^r=\{\tau_i^r,i\in \mathbb{Z}^+\}$ where $\tau_i^r$ is the overall latency of the $i$-th transaction in the order phase. $T^r$ consists of five parts: the communication latency of a transaction from the client to the OSN leader denoted by a random variable $T_{c2l}^r=\{\tau_{i,c2l}^r,i\in \mathbb{Z}^+\}$ where $\tau_{i,c2l}^r$ is the $i$-th transaction's communication latency from the client to the OSN leader, the communication latency of a transaction from the OSN leader to the OSN followers denoted by a random variable $T_{l2f}^r=\{\tau_{i,l2f}^r,i\in \mathbb{Z}^+\}$ where $\tau_{i,l2f}^r$ is the $i$-th transaction's communication latency from the OSN leader to the OSN followers, the service time of a transaction in the OSN leader denoted by a random variable $T_{s}^r=\{\tau_{i,s}^r,i\in \mathbb{Z}^+\}$ where $\tau_{i,s}^r$ is the $i$-th transaction's service time, the idle time of a transaction in the ordering service denoted by a random variable $T_{idle}^r=\{\tau_{i,idle}^r,i\in \mathbb{Z}^+\}$ where $\tau_{i,idle}^r$ is the $i$-th transaction's idle time, the queue latency of a transaction in the ordering service denoted by a random variable $T_{q}^r=\{\tau_{i,q}^r,i\in \mathbb{Z}^+\}$ where $\tau_{i,q}^r$ is the $i$-th transaction's queue latency. Therefore, we have the expected latency of a transaction in the order phase, where random variables $T^r=T_{c2l}^r+T_{l2f}^r+T_{s}^r+T_{idle}^r+T_{q}^r$, as follows,

\begin{equation}
    \begin{split}
        \mathop{\mathbb{E}}\left [ T^r \right ]=\mathop{\mathbb{E}}\left [ T_{c2l}^r \right ]+\mathop{\mathbb{E}}\left [ T_{l2f}^r \right ]+\mathop{\mathbb{E}}\left [ T_s^r \right ]+\mathop{\mathbb{E}}\left [ T_{idle}^r \right ]+\mathop{\mathbb{E}}\left [ T_q^r \right ]\;\;\;\;\;\;\;\;\;\;\\
        = \alpha+\frac{m\lambda^r}{\beta_{c2l}^r} +\alpha+\frac{(n-1)m\lambda^r}{\beta_{l2f}^r}+\frac{1}{\mu^r}\;\;\;\;\;\;\;\;\;\;\;\;\;\;\;\;\;\;\;\;\;\;\;\;\;\;\;\,\\
        +min\left ( \frac{BatchTimeout}{2}, \frac{\Delta t}{2} \right )\;\;\;\;\;\;\;\;\;\;\;\;\;\;\;\;\;\;\;\;\;\;\;\;\;\;\;\;\;\;\;\,\\
        +\left ( \frac{\rho^r}{1-\rho^r} \right )\left ( \frac{(CV_a^r)^2+(CV_s^r)^2}{2} \right )\frac{1}{\mu^r}\;\;\;\;\;\;\;\;\;\;\;\;\;\;\;\;\;\;\;
    \end{split}
\end{equation}

\begin{equation}
    \underset{min~\Delta t}{\lambda^r \Delta t}\geq BatchSize
\end{equation}

\noindent where, first, transactions between the clients and the OSN leader are transferred over the network in pipeline parallelism. Thus, we model the expected communication latency from the client to the OSN leader as $\mathop{\mathbb{E}}\left [ T_{c2f}^r \right ]=\alpha+\frac{m\lambda^r}{\beta_{c2f}^r}$, where $\lambda^r$ is the transaction arrival rate, $\beta_{c2f}^r$ is the effective bandwidth from the client to the OSN leader. Second, transactions between the OSN leader and the $k-1$ OSN followers are broadcast over the network in pipeline parallelism. Thus, we model the expected communication latency from the OSN leader to the $k-1$ OSN followers are $\mathop{\mathbb{E}}\left [ T_{l2f}^r \right ]=\alpha+\frac{(k-1)m\lambda^r}{\beta_{l2f}^r}$, where $\lambda^r$ is the given transaction arrival rate, $\beta_{l2f}^r$ is the effective bandwidth from the OSN leader to the $n-1$ OSN followers. Third, the expected service time of the OSN leader can be modeled as $\mathop{\mathbb{E}} \left [ T_s^r \right ]=\frac{1}{\mu^r}$, where $\mu^r=max(\psi^r)$. Fourth, the idle time of a transaction $T_{idle}^r$ has two cases. In case one, the \textit{BatchTimeout} condition is satisfied, and the expected idle time of a transaction is modeled as $\mathop{\mathbb{E}}\left [ T_{idle}^r \right ] = \frac{BatchTimeout}{2}$. In case two, the \textit{BatchSize} condition is satisfied, and the expected idle time of a transaction is modeled as $\mathop{\mathbb{E}}\left [ T_{idle}^r \right ] = \frac{\Delta t}{2}$, where $\Delta t$ satisfies the condition that during a period of $\Delta t$, the number of arriving transactions with a transaction arrival rate $\lambda^r$ equals to \textit{BatchSize}, meaning that the second condition is satisfied first. Fifth, the process of transaction ordering in the OSN leader is modeled as a $G/G/1$ queue. We approximate \cite{de2022latency, chen2018analytical} the $G/G/1$ queue because there is no closed-form solution of the $G/G/1$ queue \cite{zhong2018bottleneck}. Therefore, the expected queue latency of the ordering service can be modeled as $\mathop{\mathbb{E}}\left [ T_q^r \right ]=\left ( \frac{\rho^r}{1-\rho^r} \right )\left ( \frac{(CV_a^r)^2+(CV_s^r)^2}{2} \right )\frac{1}{\mu^r}$, where $\mu^r$ is the service rate of transaction ordering in the OSN leader, $CV_a^r$ is the coefficient of variation for the arrival interval in the OSN leader, and $CV_s^r$ is the coefficient of variation for the service time in the OSN leader.  

\subsection{The Validate Phase} 

\textit{Steady-State Throughput of a Committing Peer.} Assume that the transaction arrival of the validate phase follows a General distribution and the service time also follows a General distribution. We have the transaction arrival rate of the validate phase $\lambda^v=\psi^r$ because the output of the order phase becomes the input of the validate phase. Let's consider a committing peer in the validate phase as an anchor peer since there is no extra hop between the ordering service and the committing peer. Denote the throughput of a committing peer by $\psi^v$. We model the process of a committing peer in the validate phase as a $G/G/1$ queue because the validation of a block depends on the committed result of its preceding block in order to prevent double-spending attacks. It implies that the service times of the validate phase are not parallelized. Therefore, we have the steady-state throughput of a committing peer, where $\rho^v=\frac{\lambda^v}{\mu^v}$, as follows,

\begin{equation}
    \psi^v=\mu^v \rho^v=\frac{R_{disk} \cdot d}{m}\rho^v
\end{equation}

\noindent where $d$ is the written data size per disk IO and $R_{disk}$ is the number of IOs per second. For an HDD disk, the number of IOs per second is $\frac{1}{R_{disk}}=W_{seek} + 1/IOPS$, where $W_{seek}$ represents the seek latency per IO. For an SSD disk, the number of IOs per second is $\frac{1}{R_{disk}}=1/IOPS$ with no seek latency per IO.

\textit{Expected Latency in the Validate Phase.} Consider that the overall latency of a transaction in the validate phase is a random variable $T^v=\{\tau_i^v,i\in \mathbb{Z}^+\}$ where $\tau_i^v$ is the overall latency of the $i$-th transaction in the validate phase. $T^v$ consists of three parts: the communication latency of a transaction from the ordering service to the committing peer denoted by a random variable $T_{comm}^v=\{\tau_{i,comm}^v,i\in \mathbb{Z}^+\}$ where $\tau_{i,comm}^v$ is the $i$-th transaction's communication latency, the service time of a transaction in the validate phase denoted by a random variable $T_{s}^v=\{\tau_{i,s}^v,i\in \mathbb{Z}^+\}$ where $\tau_{i,s}^v$ is the $i$-th transaction's service time, the queue latency of a transaction in the validate phase denoted by a random variable $T_{q}^v=\{\tau_{i,q}^v,i\in \mathbb{Z}^+\}$ where $\tau_{i,q}^v$ is the $i$-th transaction's queue latency. Therefore, we have the expected latency of a transaction in the validate phase, where random variables $T^v=T_{comm}^v+T_{s}^v+T_{q}^v$, as follows,

\begin{equation}
    \begin{split}
        \mathop{\mathbb{E}} \left [ T^v \right ]=\mathop{\mathbb{E}} \left [ T_{comm}^v \right ] + \mathop{\mathbb{E}} \left [ T_{s}^v \right ] + \mathop{\mathbb{E}} \left [ T_{q}^v \right ]\;\;\;\;\;\;\;\;\;\;\;\;\;\;\;\;\\
        =\alpha+\frac{m\lambda^v}{\beta^v}+\frac{1}{\mu^v}+\left ( \frac{\rho^v}{1-\rho^v} \right )\left ( \frac{(CV_{a}^v)^2+(CV_{s}^v)^2}{2} \right )\frac{1}{\mu^v}
    \end{split}
\end{equation}

\noindent where, first, transactions between the ordering service and the committing peer are transferred over the network in pipeline parallelism. Thus, we model the expected communication latency from the ordering service to the committing peer as $\mathop{\mathbb{E}} \left [ T_{comm}^v \right ]=\alpha+\frac{m\lambda^v}{\beta^v} $, where $\lambda^v$ is the transaction arrival rate, $\beta^v$ is the effective bandwidth from the ordering service to the committing peer. Second, the expected service time of the validate phase can be modeled as $\mathop{\mathbb{E}} \left [ T_s^v \right ]=\frac{1}{\mu^v}$, where $\mu^v=max(\psi^v)$. Third, the expected queue latency of the validate phase can be modeled as $\mathop{\mathbb{E}}\left [ T_q^v \right ]=\left ( \frac{\rho^v}{1-\rho^v} \right )\left ( \frac{(CV_{a}^v)^2+(CV_{s}^v)^2}{2} \right )\frac{1}{\mu^v}$, where $CV_a^v$ is the coefficient of variation of arrival interval in the committing peer, and $CV_s^v$ is the coefficient of variation for service time in the committing peer. 

\section{The Measurement Framework}
This section will present a measurement method and framework to decompose and study each phase's characteristics. For a better understanding, we will use a realization to observe the random variables of each phase in Hyperledger Fabric.

\textit{Separation of Hyperledger Fabric Service and Client.} Hyperledger Fabric system can be divided into two parts: service and client. The service side is the core of the Hyperledger Fabric system and involves several components such as peer, orderer, and smart contract. The client side is a lightweight member of the Hyperledger Fabric system. Say a client of a specific peer only needs to own a membership certificate issued by Fabric CA. The relationship between the service and client sides is that each peer on the service side may have several clients while each client belongs to a specific peer. The client and service sides take different jobs. For example, the client side generates new transaction proposals for the endorsing peer and new transactions for the ordering service. In contrast, the service side handles these new transaction proposals and transactions from the client side. People often separate the client from the service side for most real-world applications because most transactions come from the client side, not the service side. How does the client side communicate with the service side? Hyperledger Fabric 2.2 LTS provides a toolkit named Fabric SDK Nodejs that enables many essential functions between the client and service sides. For example, the Fabric SDK Nodejs allows the client to prepare and submit a new transaction proposal (or/and a transaction) to the service side. Upon the transaction proposal endorsed by the endorsing peer or a block is committed by a peer, the service side will notify the client side via a callback function. Therefore, based on the functions provided by the Fabric SDK Nodejs, our measurement framework can trace the transaction flow between the client and service sides.

\begin{figure}[t]
  \centering
  \includegraphics[scale=0.76]{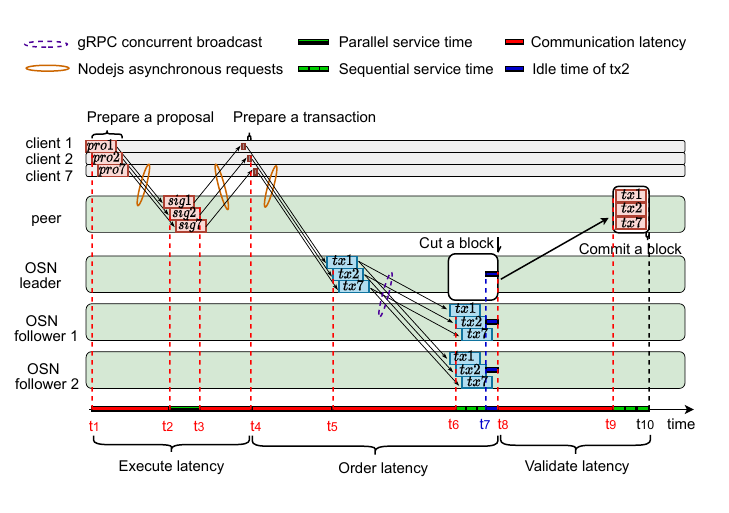}
  \caption{A measurement framework decomposes each phase in Hyperledger Fabric from a transaction-based perspective, where we use a realization (i.e., transaction $tx2$) to observe the random variables, e.g., $T^e$, $T^r$, $T^v$, and identify each phase's start and end time of each phase. Specifically, the execute phase starts at timestamp $t_1$, at which the client prepares a transaction proposal and submits it to the endorsing peer. The order phase starts at timestamp $t_4$, where the client prepares a transaction for the order phase. The validate phase starts at timestamp $t_8$, at which a new block is created. Finally, the validate phase ends at timestamp $t_{10}$, at which the committing peer successfully commits the block.} 
\end{figure}

\textit{Container-based Resource Management.} Most related works about Hyperledger Fabric service are deployed in a container-based isolated environment because it is convenient for resource management, and the cost of the docker container can be neglected. Our measurement framework provides a method of dominant resource identification for each phase in Hyperledger Fabric via container-based resource management. First, we use the command of the cpuset supplied by docker-compose~v2.4 and control the number of available CPU cores for an endorsing peer in the execute phase. Second, we use the Linux iftop to measure the network efficient bandwidth of end-to-end communication. Third, we use the command of device\_write\_iops provided by docker-compose~v2.4 and control the disk IOs of a committing peer in the validate phase. Fourth, we apply the network time protocol (NTP) among all hosts (i.e., containers) for distributed system clock synchronization. And NTP achieves an error of around 2 to 3~miliseconds among all hosts within a cluster (i.e., via connecting to a single standard time server's IP address in practice), which is negligible compared to a transaction's latency.   

\textit{Steady Requests via Asynchronous \& Concurrent Threads.} There are two requirements for arrival transactions in Hyperledger Fabric. The first requirement is to generate a large number of steady arriving transactions from the client side to utilize the resource on the service side of Hyperledger Fabric. In practice, a single client usually cannot generate enough arriving transactions to utilize a peer highly; thus, we present a measurement framework with multiple clients. For each client, it is implemented based on the Fabric SDK Nodejs. For each request, i.e., a Nodejs thread, Fabric SDK Nodejs will launch an asynchronous thread to handle this request due to the underlying mechanism of Nodejs. It implies that the running threads are asynchronous and that no two threads need to wait for each other. For each Nodejs request thread inside, there is a blocking function. It is because a request must go through all the phases individually and wait until the peer notifies the commit status. Each client can fully use multiple CPU cores in a single machine. Therefore we can satisfy the first requirement and stably generate many arrival transactions by launching multiple clients on multiple machines, where each machine hosts one client. The second requirement is to control the transaction arrival rate of the first phase, i.e., the execute phase in Hyperledger Fabric. We implement the Poisson-based arriving transactions for the execute phase by introducing an exponentially distributed inter-arrival interval to each request. And the overall arriving transactions from the clients also follow a Poisson distribution due to the merging property of Poisson distribution. The transaction arrival rate is achieved by adding a specific exponentially-distributed latency to each request. And in this way, the second requirement of the measurement framework is satisfied, and a particular transaction arrival rate of the first phase in Hyperledger Fabric is achieved.

\begin{figure}[t]
  \centering
  \includegraphics[scale=0.74]{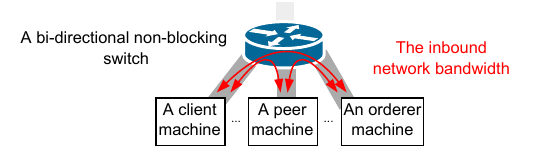}
  \caption{A network topology where all machines in a cluster are connected via a bi-directional non-blocking switch.}
\end{figure}

\textit{Performance Metrics in Each Phase.} An overall transaction flow consists of three phases: the execute phase, the order phase, and the validate phase. Existing works lack a timestamp scheme to identify each phase in Hyperledger Fabric. In this work, we present a timestamp scheme to identify each phase’s start and end times from a transaction-based perspective, as shown in Fig.~4. Specifically, the execute phase of transaction $tx2$ starts at timestamp $t_1$ when the client prepares the transaction proposal and submits it to the endorsing peer. The order phase of transaction $tx2$ starts (i.e., the execute phase of transaction $tx2$ ends) at timestamp $t_4$ when the client submits the transaction to the endorsing service. The validate phase of transaction $tx2$ starts (i.e., the order phase of transaction $tx2$ ends) at timestamp $t_8$ when a new block is created. Finally, the validate phase of transaction $tx2$ ends at timestamp $t_{10}$ when the new block is successfully committed. Now, we will take transaction $tx2$ as an example and analyze the performance metric of each phase from a transaction-based perspective.

In the execute phase, the client submits transaction proposal~$tx2$ to the endorsing peer at timestamp~$t_1$. The endorsing peer receives the transaction proposal at timestamp~$t_2$ and replies to the client at timestamp~$t_3$. Finally, the client receives the reply at timestamp~$t_4$. Therefore, the expected communication latency of a transaction spent between the client and the endorsing peer $\mathop{\mathbb{E}}\left [ T_{comm}^e \right ]$, and the expected latency of a transaction spent in the execute phase $\mathop{\mathbb{E}}\left [ T^e \right ]$ are, as follows,

\begin{equation}
    \mathop{\mathbb{E}}\left [ T_{comm}^e \right ]=\mathop{\mathbb{E}} \left [ t_2-t_1 +t_4-t_3\right ]; \mathop{\mathbb{E}}\left [ T^e \right ]=\mathop{\mathbb{E}} \left [ t_4-t_1 \right ]
\end{equation}

In the order phase, the client submits transaction $tx2$ to the OSN leader at timestamp $t_4$. The OSN leader receives the transaction at timestamp $t_5$ and then broadcasts the transaction $tx2$ to $k-1$ OSN followers. And a consensus among $k$ OSNs on transaction $tx2$ is achieved at timestamp $t_6$. But the \textit{BatchSize} condition is not satisfied yet, and transaction $tx2$ waits for the block cutter for $t_8-t_7$ seconds until a new block is created by the OSN leader at timestamp $t_8$. Therefore, the expected communication latency of a transaction spent from the client to the OSN leader $\mathop{\mathbb{E}}\left [ T_{c2l}^r \right ]$, the expected communication latency of a transaction spent from the OSN leader to the $k-1$ OSN followers $\mathop{\mathbb{E}}\left [ T_{l2f}^r \right ]$, and the expected latency of a transaction spent in the order phase $\mathop{\mathbb{E}}\left [ T^r \right ]$ are, as follows,

\begin{equation}
    \begin{split}
        \mathop{\mathbb{E}}\left [ T_{c2l}^r \right ]=\mathop{\mathbb{E}} \left [ t_5-t_4\right ]; \mathop{\mathbb{E}}\left [ T_{l2f}^r \right ]=\mathop{\mathbb{E}} \left [ t_6-t_5\right ];\\ \mathop{\mathbb{E}}\left [ T^r \right ]=\mathop{\mathbb{E}} \left [ t_8-t_4 \right ]\;\;\;\;\;\;\;\;\;\;\;\;\;\;\;\;\;\;
    \end{split}
\end{equation}

In the validate phase, the ordering service creates and sends a new block at timestamp $t_8$, and the committing peer in the validate phase receives the block at timestamp $t_9$. And the block is committed in the validate phase at timestamp $t_{10}$. Therefore, the expected communication latency of a transaction spent from the ordering service to the committing peer in the validate phase $\mathop{\mathbb{E}}\left [ T_{comm}^v \right ]$, and the expected latency of a transaction spent in the validate phase $\mathop{\mathbb{E}}\left [ T^v \right ]$ are, as follows,

\begin{equation}
    \mathop{\mathbb{E}}\left [ T_{comm}^v \right ]=\mathop{\mathbb{E}} \left [ t_9-t_8\right ]; \mathop{\mathbb{E}}\left [ T^v \right ]=\mathop{\mathbb{E}} \left [ t_{10}-t_8 \right ]
\end{equation}

\begin{table}[t]
\footnotesize 
	\centering
	\caption{Experiment Environments}
	\scalebox{0.90}{
\begin{tabular}{|c|l|l|}
\hline  
Configuration     & Cluster 1 (Local)    & Cluster 2 (Alibaba Cloud)          \\ \hline \hline
CPU Model         & Intel i7 3.4 GHz & Intel Platinum 8269 CY 3.2 GHz \\ \hline
Memory Size       & 8 GB             & 16 GB                          \\ \hline
Disk Type         & HDD        & SSD                     \\ \hline
Inbound Bandwidth & 1 Gbps           & 10 Gbps                        \\ \hline
\end{tabular}
}
\end{table}

\begin{figure*}[t]
	\centering
	\subfigure[The effects of cpu cores on decomposed latencies $T^e_{comm}$, $T^e_{s}$, and $T^e_{q}$ in the execute phase]{%
		\includegraphics[width=0.810\textwidth]{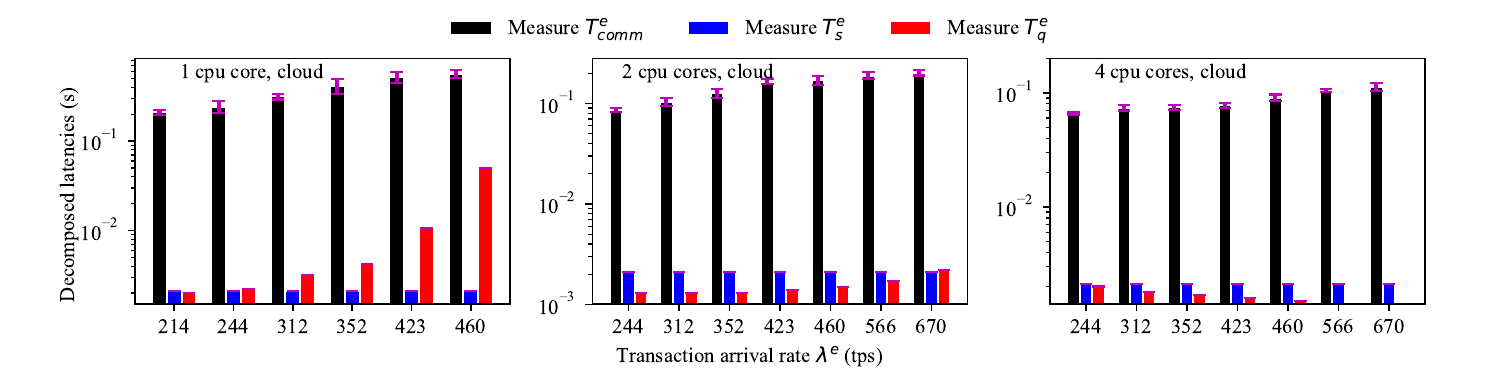}%
	}
	
	\subfigure[The effects of cpu cores on latency $T^e$ in the execute phase]{%
		\includegraphics[width=0.810\textwidth]{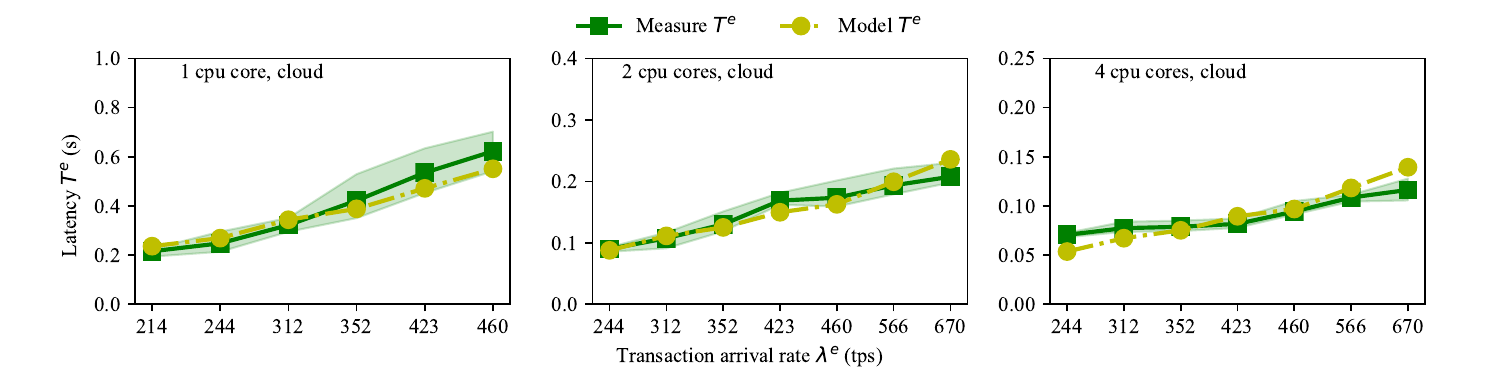}%
	}
 
	\subfigure[The effects of cpu cores on bandwidth utilization $\beta^e$ in the execute phase]{%
		\includegraphics[width=0.810\textwidth]{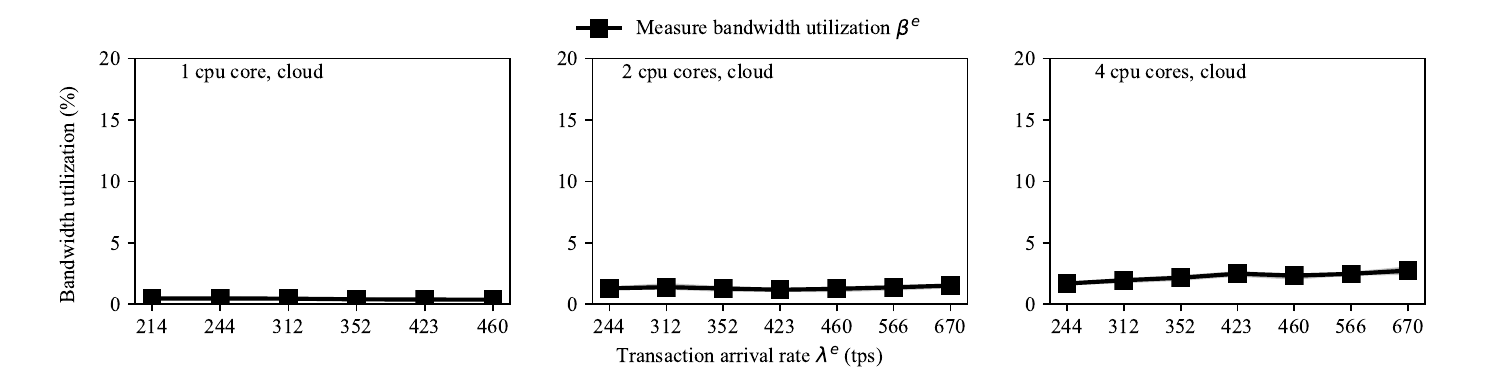}%
	}
  
        \caption{The effects of CPU cores on the latency and bandwidth utilization in the execute phase. There is an endorsing peer with $c=1, 2, 4$ CPU core(s), respectively. The rest machines are seven clients and three ordering service nodes. All machines are connected via a bi-directional non-blocking switch with a cloud Ethernet network of 10 Gbps. Remark that  $T^e=T^e_{comm}+T^e_{s}+T^e_{q}$.}
\end{figure*}

\begin{figure*}[t]
	\centering
        \subfigure[The effects of OSNs on decomposed latencies $T^r_{c2l}$, $T^r_{l2f}$, $T^r_{idle}$, $T^r_{s}$, and $T^r_{q}$ in the order phase]{%
        \includegraphics[width=0.810\textwidth]{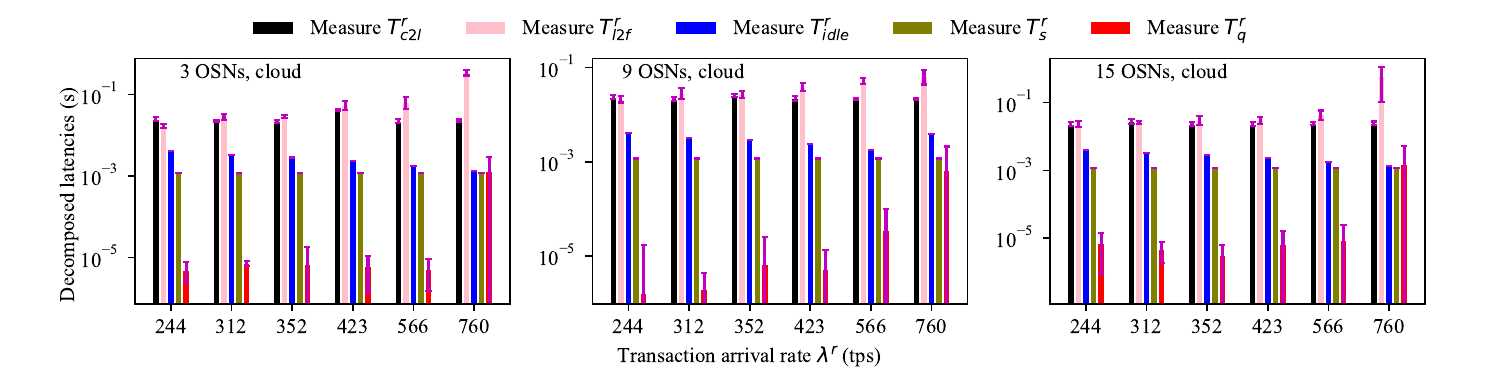}%
	}
	
        \subfigure[The effects of OSNs on latency $T^r$ in the order phase]{%
        \includegraphics[width=0.810\textwidth]{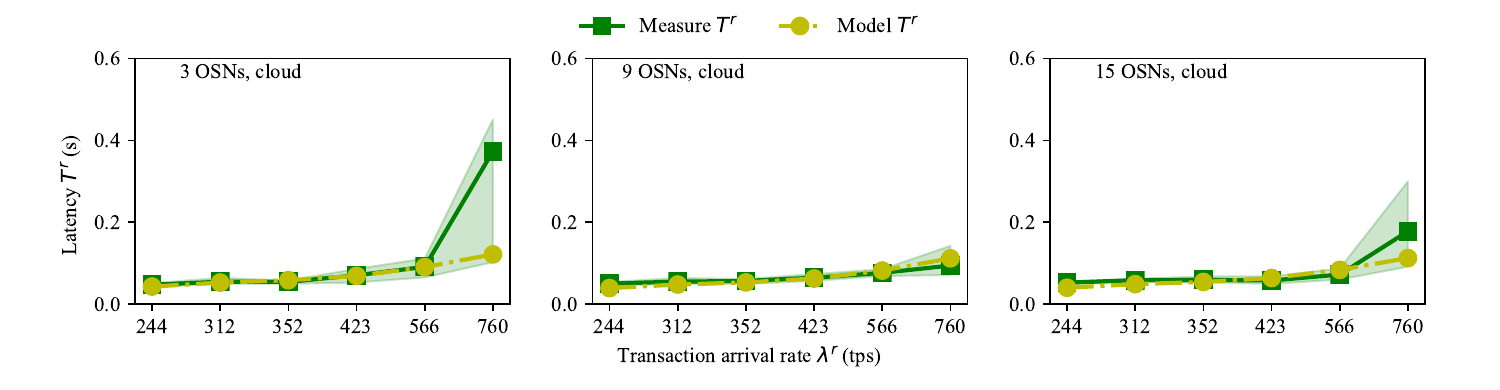}%
	}
 
        \subfigure[The effects of OSNs on bandwidth utilization in the order phase]{%
        \includegraphics[width=0.810\textwidth]{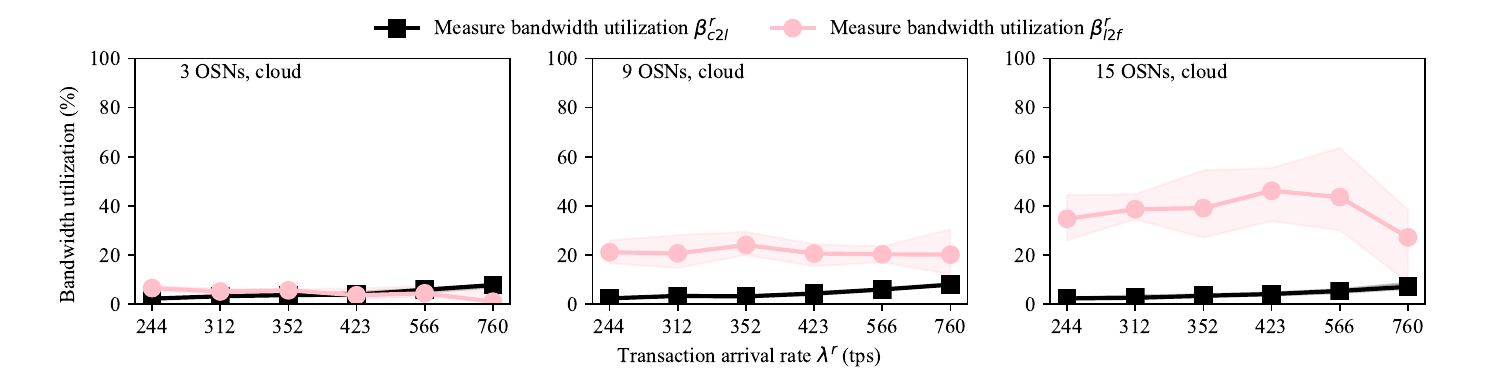}%
	}
  
        \caption{The effects of OSNs on the latency and bandwidth utilization in the order phase. There is an ordering service with $k=3, 9, 15$ OSNs, respectively. The \textit{BatchSize} is 2 and the \textit{BatchTimeout} is 1. The rest machines are seven clients and one peer. All machines are connected via a bi-directional non-blocking switch with a cloud Ethernet network of 10 Gbps. Remark that $T^r=T^r_{c2l}+T^r_{l2f}+T^r_{idle}+T^r_{s}+T^e_{q}$.}
\end{figure*}

\begin{figure*}[t]
	\centering
        \subfigure[The effects of storage devices on decomposed latencies $T^v_{comm}$, $T^r_{s}$, and $T^r_{q}$ in the validate phase]{%
        \includegraphics[width=0.893\textwidth]{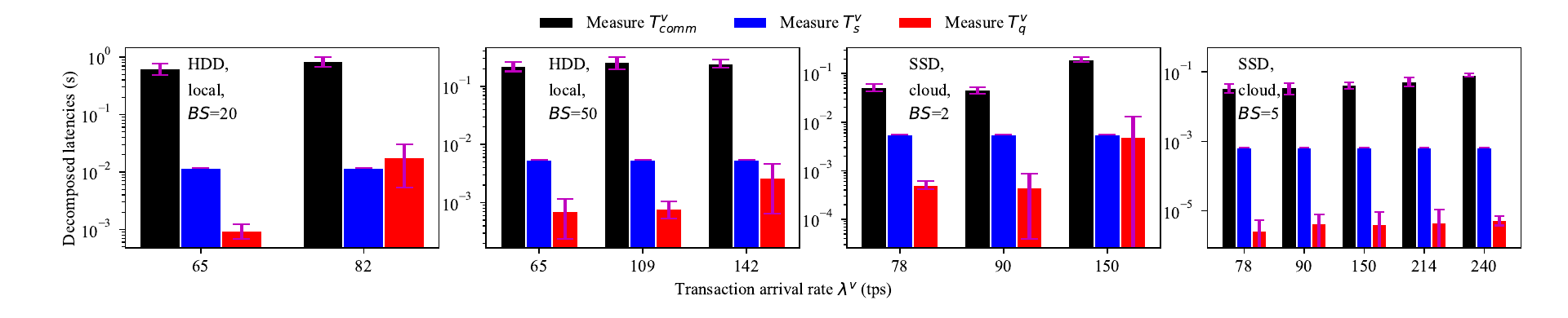}%
	}
	
        \subfigure[The effects of storage devices on latency $T^v$ in the validate phase]{%
        \includegraphics[width=0.893\textwidth]{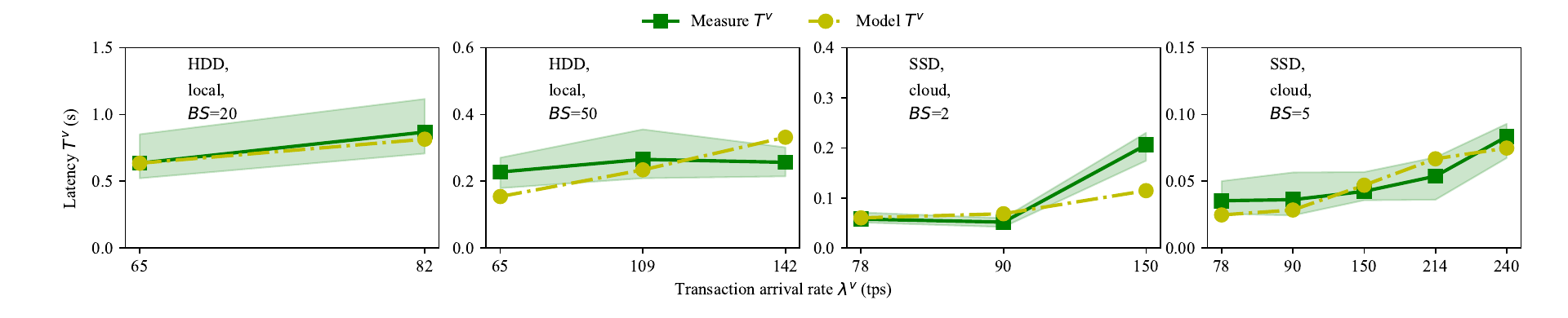}%
	}
 
        \subfigure[The effects of storage devices on bandwidth utilization in the validate phase]{%
        \includegraphics[width=0.893\textwidth]{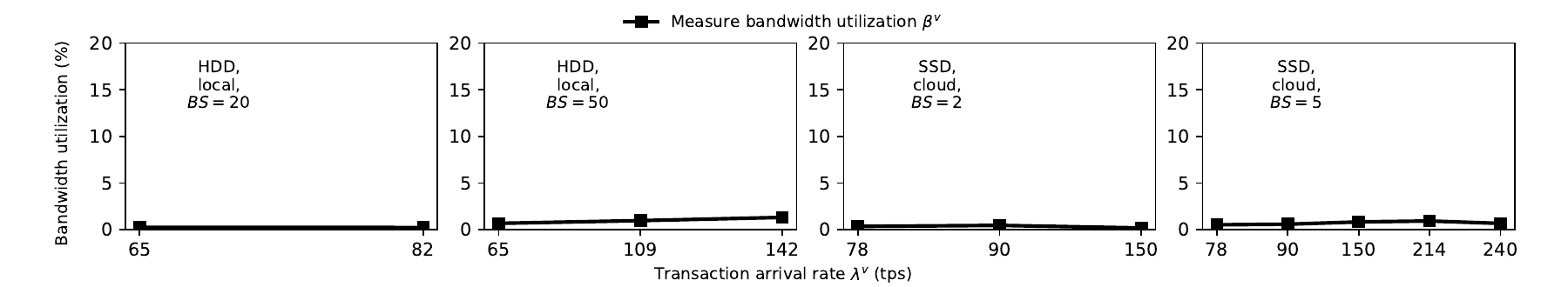}
	}
  
        \caption{The effects of storage devices on the latency and bandwidth utilization in the validate phase. There is a committing peer has two different storage devices, i.e., HDD and SSD, respectively. The rest machines are seven clients and three ordering service nodes. All machines are connected via a bi-directional non-blocking switch. Remark that $T^v=T^v_{comm}+T^v_{s}+T^v_{q}$.}
\end{figure*}

\section{Validation Results}
This section will validate the model results of each phase in Hyperledger Fabric.

\textit{Experiment Environments.} TABLE~2 shows the experiment environments. The proposed performance model will be validated in a local cluster and an Alibaba cloud cluster. Specifically, in the local cluster, we assign seven machines for seven clients, one machine for one peer, and fifteen machines for fifteen ordering service nodes. Each machine hosts a 4-core Intel i7 CPU of 3.40~GHz, 8~GB memory, and a 1~TB hard disk. All machines are connected to a 1~Gbps Ethernet network. In the Alibaba cloud cluster, we assign seven machines for seven clients, one machine for one peer, and fifteen machines for fifteen ordering service nodes. Each machine hosts an Intel Platinum 8269~CY 8-vCPU of 3.2 GHz, 16~GB memory, and a 40~GB SSD disk. All machines are connected to a 10~Gbps Ethernet network. Finally, the network topology of a cluster is shown in Fig.~5, where all machines are directly connected via a bi-directional non-blocking switch.

\subsection{The Execute Phase}

Fig.~6 shows the effects of CPU cores on the latency and bandwidth utilization in the execute phase, where an endorsing peer has $c=1, 2, 4$ CPU core(s), respectively. Fig.~6~(a) shows the effects of CPU cores on the decomposed latencies $T_{comm}^e$, $T_s^e$, $T_q^e$ in the execute phase. Fig.~6~(b) shows the effects of CPU cores on the latency $T^e$ in the execute phase. Fig.~6~(c) shows the effects of CPU cores on the bandwidth utilization in the execute phase. 


In summary, first, the experiment results validate that an endorsing peer in the execute phase has scalable throughput with the number of CPU cores due to parallel service times among CPU cores in an endorsing peer. Second, it shows that the communication latency dominates the transaction latency spent in the execute phase. 

\subsection{The Order Phase}

Fig.~7 shows the effects of OSNs on the latency and bandwidth utilization in the order phase, where an ordering service has $k=3, 9, 15$ ordering service nodes, respectively. Fig.~7~(a) shows the effects of OSNs on the decomposed latencies $T_{c2l}^r$, $T_{l2f}^r$, $T_{idle}^r$, $T_s^r$, $T_q^r$ in the order phase. Fig.~7~(b) shows the effects of OSNs on the latency $T^r$ in the order phase. Fig.~7~(c) shows the effects of OSNs on bandwidth utilization in the order phase. 


In summary, first, the result validates that a larger batch size leads to a more significant throughput of the OSN leader in the ordering service, because a proper larger batch size better utilizes network bandwidth and hence gains low communication latency in the order phase. Second, the result validates that the Raft-based ordering service shows good scalability with the number of ordering service nodes due to concurrent processing between the OSN leader and its followers. Third, it shows that the communication latency, in particular the communication latency of the Raft protocol, dominates the overall latency of a transaction spent in the order phase in both 1~Gbit/s and 10~Gbit/s Ethernet networks.

\subsection{The Validate Phase}

Fig.~8 shows the effects of storage devices on the latency and bandwidth utilization in the validate phase, where a committing peer has two different storage devices, i.e., HDD, SSD, respectively. In a local cluster of 1 Gbps network, the throughput of a committing peer with an HDD and a \textit{BatchSize} of 20 is 85~transactions per second. The throughput of a committing peer with an HDD and a \textit{BatchSize} of 50 is 185 transactions per second. In a cloud cluster of 10~Gbps network, the throughput of a committing peer with an SSD and a \textit{BatchSize} of 2 is 180~transactions per second. The throughput of a committing peer with an SSD and a \textit{BatchSize} of 5 is 1490~transactions per second. 


In summary, first, the result validates that a larger batch size leads to a larger throughput of the committing peer in the validate phase, because a larger number of data can be written to disk per IO. Second, the result validates that SSD shows higher throughput than HDD in the validate phase, due to smaller service time per block in SSD, in particular blocks are committed sequentially. Third, it shows that the communication latency dominates the overall latency of a transaction spent in the validate phase in both 1 Gbit/s and 10 Gbit/s Ethernet networks.

\section{Conclusion}

In conclusion, this paper presents a measurement framework to characterize each phase’s transaction and block data in Hyperledger Fabric based on the Fabric SDK Nodejs, where we thoroughly analyze and open-source the implementation details of the measurement framework. We then propose a stochastic computation model for each phase's performance analysis and an alpha-beta communication model to analyze the corresponding communication latency. Specifically, in the execute phase, the result shows that an endorsing peer in the execute phase has scalable throughput with the number of CPU cores due to parallel service times among CPU cores in an endorsing peer. In the order phase, the result shows that a larger batch size leads to a more significant throughput of the OSN leader in the ordering service because a proper larger batch size better utilizes network bandwidth and hence gains low communication latency in the order phase. And the Raft-based ordering service shows good scalability with the number of ordering service nodes due to concurrent processing between the OSN leader and its followers. Moreover, the communication latency, in particular the communication latency of the Raft protocol, dominates the overall latency of a transaction spent in the order phase in both 1 Gbit/s and 10 Gbit/s Ethernet networks. In the validate phase, the result shows that a larger batch size leads to a larger throughput of the committing peer in the validate phase because a larger number of data can be written to disk per IO. And SSD shows higher throughput than HDD in the validate phase, due to smaller service time per block in SSD, particularly blocks are committed sequentially. We validate the proposed performance model on both local and cloud clusters. The results and the performance model help guide the resource allocation of Hyperledger Fabric service.

\ifCLASSOPTIONcaptionsoff
  \newpage
\fi



%



\bibliographystyle{IEEEtran}
\bibliography{IeeeTnseHPArxivV12}

\begin{thebibliography}{10}
\providecommand{\url}[1]{#1}
\csname url@samestyle\endcsname
\providecommand{\newblock}{\relax}
\providecommand{\bibinfo}[2]{#2}
\providecommand{\BIBentrySTDinterwordspacing}{\spaceskip=0pt\relax}
\providecommand{\BIBentryALTinterwordstretchfactor}{4}
\providecommand{\BIBentryALTinterwordspacing}{\spaceskip=\fontdimen2\font plus
\BIBentryALTinterwordstretchfactor\fontdimen3\font minus \fontdimen4\font\relax}
\providecommand{\BIBforeignlanguage}[2]{{%
\expandafter\ifx\csname l@#1\endcsname\relax
\typeout{** WARNING: IEEEtran.bst: No hyphenation pattern has been}%
\typeout{** loaded for the language `#1'. Using the pattern for}%
\typeout{** the default language instead.}%
\else
\language=\csname l@#1\endcsname
\fi
#2}}
\providecommand{\BIBdecl}{\relax}
\BIBdecl

\bibitem{nakamoto2008bitcoin}
S.~Nakamoto, ``Bitcoin: A peer-to-peer electronic cash system,'' \emph{Decentralized Business Review}, p. 21260, 2008.

\bibitem{wang2022dissecting}
C.~Wang, X.~Chu, and Y.~Qin, ``Dissecting mining pools of bitcoin network: Measurement, analysis and modeling,'' \emph{IEEE Transactions on Network Science and Engineering}, 2022.

\bibitem{gai2019permissioned}
K.~Gai, Y.~Wu, L.~Zhu, L.~Xu, and Y.~Zhang, ``Permissioned blockchain and edge computing empowered privacy-preserving smart grid networks,'' \emph{IEEE Internet of Things Journal}, vol.~6, no.~5, pp. 7992--8004, 2019.

\bibitem{putra2021trust}
G.~D. Putra, V.~Dedeoglu, S.~S. Kanhere, R.~Jurdak, and A.~Ignjatovic, ``Trust-based blockchain authorization for iot,'' \emph{IEEE Transactions on Network and Service Management}, vol.~18, no.~2, pp. 1646--1658, 2021.

\bibitem{helliar2020permissionless}
C.~V. Helliar, L.~Crawford, L.~Rocca, C.~Teodori, and M.~Veneziani, ``Permissionless and permissioned blockchain diffusion,'' \emph{International Journal of Information Management}, vol.~54, p. 102136, 2020.

\bibitem{bakos2021permissioned}
Y.~Bakos, H.~Halaburda, and C.~Mueller-Bloch, ``When permissioned blockchains deliver more decentralization than permissionless,'' \emph{Communications of the ACM}, vol.~64, no.~2, pp. 20--22, 2021.

\bibitem{peng2021privacy}
L.~Peng, W.~Feng, Z.~Yan, Y.~Li, X.~Zhou, and S.~Shimizu, ``Privacy preservation in permissionless blockchain: A survey,'' \emph{Digital Communications and Networks}, vol.~7, no.~3, pp. 295--307, 2021.

\bibitem{de2020survey}
E.~J. De~Aguiar, B.~S. Fai{\c{c}}al, B.~Krishnamachari, and J.~Ueyama, ``A survey of blockchain-based strategies for healthcare,'' \emph{ACM Computing Surveys (CSUR)}, vol.~53, no.~2, pp. 1--27, 2020.

\bibitem{lao2020survey}
L.~Lao, Z.~Li, S.~Hou, B.~Xiao, S.~Guo, and Y.~Yang, ``A survey of iot applications in blockchain systems: Architecture, consensus, and traffic modeling,'' \emph{ACM Computing Surveys (CSUR)}, vol.~53, no.~1, pp. 1--32, 2020.

\bibitem{lakhan2022blockchain}
A.~Lakhan, M.~A. Mohammed, J.~Nedoma, R.~Martinek, P.~Tiwari, and N.~Kumar, ``Blockchain-enabled cybersecurity efficient iioht cyber-physical system for medical applications,'' \emph{IEEE Transactions on Network Science and Engineering}, 2022.

\bibitem{peng2023patient}
G.~Peng, A.~Zhang, and X.~Lin, ``Patient-centric fine-grained access control for electronic medical record sharing with security via dual-blockchain,'' \emph{IEEE Transactions on Network Science and Engineering}, 2023.

\bibitem{androulaki2018hyperledger}
E.~Androulaki, A.~Barger, V.~Bortnikov, C.~Cachin, K.~Christidis, A.~De~Caro, D.~Enyeart, C.~Ferris, G.~Laventman, Y.~Manevich \emph{et~al.}, ``Hyperledger fabric: a distributed operating system for permissioned blockchains,'' in \emph{Proceedings of the thirteenth EuroSys conference}, 2018, pp. 1--15.

\bibitem{klenik2022porting}
A.~Klenik and I.~Kocsis, ``Porting a benchmark with a classic workload to blockchain: Tpc-c on hyperledger fabric,'' in \emph{Proceedings of the 37th ACM/SIGAPP Symposium on Applied Computing}, 2022, pp. 290--298.

\bibitem{jeong2021multilateral}
J.~Jeong, D.~Kim, S.-Y. Ihm, Y.~Lee, and Y.~Son, ``Multilateral personal portfolio authentication system based on hyperledger fabric,'' \emph{ACM Transactions on Internet Technology (TOIT)}, vol.~21, no.~1, pp. 1--17, 2021.

\bibitem{lohachab2021performance}
A.~Lohachab, S.~Garg, B.~H. Kang, and M.~B. Amin, ``Performance evaluation of hyperledger fabric-enabled framework for pervasive peer-to-peer energy trading in smart cyber--physical systems,'' \emph{Future Generation Computer Systems}, vol. 118, pp. 392--416, 2021.

\bibitem{fan2021performance}
C.~Fan, S.~Ghaemi, H.~Khazaei, Y.~Chen, and P.~Musilek, ``Performance analysis of the iota dag-based distributed ledger,'' \emph{ACM Transactions on Modeling and Performance Evaluation of Computing Systems}, vol.~6, no.~3, pp. 1--20, 2021.

\bibitem{hang2021optimal}
L.~Hang and D.-H. Kim, ``Optimal blockchain network construction methodology based on analysis of configurable components for enhancing hyperledger fabric performance,'' \emph{Blockchain: Research and Applications}, vol.~2, no.~1, p. 100009, 2021.

\bibitem{wang2020performance}
C.~Wang and X.~Chu, ``Performance characterization and bottleneck analysis of hyperledger fabric,'' in \emph{2020 IEEE 40th International Conference on Distributed Computing Systems (ICDCS)}.\hskip 1em plus 0.5em minus 0.4em\relax IEEE, 2020, pp. 1281--1286.

\bibitem{barger2021byzantine}
A.~Barger, Y.~Manevich, H.~Meir, and Y.~Tock, ``A byzantine fault-tolerant consensus library for hyperledger fabric,'' in \emph{2021 IEEE International Conference on Blockchain and Cryptocurrency (ICBC)}.\hskip 1em plus 0.5em minus 0.4em\relax IEEE, 2021, pp. 1--9.

\bibitem{xu2021latency}
X.~Xu, G.~Sun, L.~Luo, H.~Cao, H.~Yu, and A.~V. Vasilakos, ``Latency performance modeling and analysis for hyperledger fabric blockchain network,'' \emph{Information Processing \& Management}, vol.~58, no.~1, p. 102436, 2021.

\bibitem{yuan2020performance}
P.~Yuan, K.~Zheng, X.~Xiong, K.~Zhang, and L.~Lei, ``Performance modeling and analysis of a hyperledger-based system using gspn,'' \emph{Computer Communications}, vol. 153, pp. 117--124, 2020.

\bibitem{meng2021consortium}
T.~Meng, Y.~Zhao, K.~Wolter, and C.-Z. Xu, ``On consortium blockchain consistency: A queueing network model approach,'' \emph{IEEE Transactions on Parallel and Distributed Systems}, vol.~32, no.~6, pp. 1369--1382, 2021.

\bibitem{desai2021blockfla}
H.~B. Desai, M.~S. Ozdayi, and M.~Kantarcioglu, ``Blockfla: Accountable federated learning via hybrid blockchain architecture,'' in \emph{Proceedings of the eleventh ACM conference on data and application security and privacy}, 2021, pp. 101--112.

\bibitem{gai2022blockchain}
K.~Gai, Y.~She, L.~Zhu, K.-K.~R. Choo, and Z.~Wan, ``A blockchain-based access control scheme for zero trust cross-organizational data sharing,'' \emph{ACM Transactions on Internet Technology (TOIT)}, 2022.

\bibitem{fotia2022trust}
L.~Fotia, F.~C. Delicato, and G.~Fortino, ``Trust in edge-based internet of things architectures: State of the art and research challenges,'' \emph{ACM Computing Surveys (CSUR)}, 2022.

\bibitem{neha2022systematic}
B.~Neha, S.~K. Panda, P.~K. Sahu, K.~S. Sahoo, and A.~H. Gandomi, ``A systematic review on osmotic computing,'' \emph{ACM Transactions on Internet of Things}, vol.~3, no.~2, pp. 1--30, 2022.

\bibitem{berendea2020fair}
N.~Berendea, H.~Mercier, E.~Onica, and E.~Riviere, ``Fair and efficient gossip in hyperledger fabric,'' in \emph{2020 IEEE 40th International Conference on Distributed Computing Systems (ICDCS)}.\hskip 1em plus 0.5em minus 0.4em\relax IEEE, 2020, pp. 190--200.

\bibitem{gorenflo2020fastfabric}
C.~Gorenflo, S.~Lee, L.~Golab, and S.~Keshav, ``Fastfabric: Scaling hyperledger fabric to 20 000 transactions per second,'' \emph{International Journal of Network Management}, vol.~30, no.~5, p. e2099, 2020.

\bibitem{sharma2019blurring}
A.~Sharma, F.~M. Schuhknecht, D.~Agrawal, and J.~Dittrich, ``Blurring the lines between blockchains and database systems: the case of hyperledger fabric,'' in \emph{Proceedings of the 2019 International Conference on Management of Data}, 2019, pp. 105--122.

\bibitem{nasir2018performance}
Q.~Nasir, I.~A. Qasse, M.~Abu~Talib, and A.~B. Nassif, ``Performance analysis of hyperledger fabric platforms,'' \emph{Security and Communication Networks}, vol. 2018, 2018.

\bibitem{nakaike2020hyperledger}
T.~Nakaike, Q.~Zhang, Y.~Ueda, T.~Inagaki, and M.~Ohara, ``Hyperledger fabric performance characterization and optimization using goleveldb benchmark,'' in \emph{2020 IEEE International Conference on Blockchain and Cryptocurrency (ICBC)}.\hskip 1em plus 0.5em minus 0.4em\relax IEEE, 2020, pp. 1--9.

\bibitem{nguyen2019impact}
T.~S.~L. Nguyen, G.~Jourjon, M.~Potop-Butucaru, and K.~L. Thai, ``Impact of network delays on hyperledger fabric,'' in \emph{IEEE INFOCOM 2019-IEEE Conference on Computer Communications Workshops (INFOCOM WKSHPS)}.\hskip 1em plus 0.5em minus 0.4em\relax IEEE, 2019, pp. 222--227.

\bibitem{kuzlu2019performance}
M.~Kuzlu, M.~Pipattanasomporn, L.~Gurses, and S.~Rahman, ``Performance analysis of a hyperledger fabric blockchain framework: throughput, latency and scalability,'' in \emph{2019 IEEE international conference on blockchain (Blockchain)}.\hskip 1em plus 0.5em minus 0.4em\relax IEEE, 2019, pp. 536--540.

\bibitem{thakkar2018performance}
P.~Thakkar, S.~Nathan, and B.~Viswanathan, ``Performance benchmarking and optimizing hyperledger fabric blockchain platform,'' in \emph{2018 IEEE 26th International Symposium on Modeling, Analysis, and Simulation of Computer and Telecommunication Systems (MASCOTS)}.\hskip 1em plus 0.5em minus 0.4em\relax IEEE, 2018, pp. 264--276.

\bibitem{thakkar2021scaling}
P.~Thakkar and S.~Natarajan, ``Scaling blockchains using pipelined execution and sparse peers,'' in \emph{Proceedings of the ACM Symposium on Cloud Computing}, 2021, pp. 489--502.

\bibitem{sukhwani2018performance}
H.~Sukhwani, N.~Wang, K.~S. Trivedi, and A.~Rindos, ``Performance modeling of hyperledger fabric (permissioned blockchain network),'' in \emph{2018 IEEE 17th International Symposium on Network Computing and Applications (NCA)}.\hskip 1em plus 0.5em minus 0.4em\relax IEEE, 2018, pp. 1--8.

\bibitem{yan2018efficient}
F.~Yan, Y.~He, O.~Ruwase, and E.~Smirni, ``Efficient deep neural network serving: Fast and furious,'' \emph{IEEE Transactions on Network and Service Management}, vol.~15, no.~1, pp. 112--126, 2018.

\bibitem{di2022performability}
M.~Di~Mauro, G.~Galatro, M.~Longo, F.~Postiglione, and M.~Tambasco, ``Performability analysis of containerized ims through queueing networks and stochastic models,'' in \emph{NOMS 2022-2022 IEEE/IFIP Network Operations and Management Symposium}.\hskip 1em plus 0.5em minus 0.4em\relax IEEE, 2022, pp. 1--8.

\bibitem{zhang2019mark}
C.~Zhang, M.~Yu, W.~Wang, and F.~Yan, ``$\{$MArk$\}$: Exploiting cloud services for $\{$Cost-Effective$\}$,$\{$SLO-Aware$\}$ machine learning inference serving,'' in \emph{2019 USENIX Annual Technical Conference (USENIX ATC 19)}, 2019, pp. 1049--1062.

\bibitem{de2022latency}
L.~De~Simone, M.~Di~Mauro, R.~Natella, and F.~Postiglione, ``A latency-driven availability assessment for multi-tenant service chains,'' \emph{IEEE Transactions on Services Computing}, 2022.

\bibitem{chen2018analytical}
N.~Chen, X.~Xie, P.~A. Bain, M.~P. Mundt, L.~Zheng, and J.~Li, ``An analytical framework for modeling, analysis, and improvement of team communication and collaboration process in primary care clinics,'' \emph{IEEE Transactions on Automation Science and Engineering}, vol.~16, no.~3, pp. 1148--1162, 2018.

\bibitem{zhong2018bottleneck}
X.~Zhong, A.~M. Prakash, L.~Petty, and R.~A. James, ``Bottleneck analysis to reduce primary care to specialty care referral delay,'' \emph{IEEE Transactions on Automation Science and Engineering}, vol.~16, no.~1, pp. 61--73, 2018.

\end{thebibliography}

\end{document}